\documentclass[pra,aps,twocolumn,superscriptaddress,showpacs,tightenlines,amsmath]{revtex4}
\usepackage{epsfig}
\usepackage{amsmath,amssymb}

\begin{document}

\title{Quantum control on entangled bipartite qubits}

\author{Francisco Delgado}
\email{fdelgado@itesm.mx}
\affiliation{Mathematics and Physics Department, Quantum Information Processing Group, Tecnologico de Monterrey, Campus Estado de Mexico, Atizapan, Estado de Mexico, CP. 52926, Mexico}

\date{\today}

\begin{abstract}
Ising interaction between qubits could produce distortion in entangled pairs generated for engineering purposes (as in quantum computation) in presence of parasite magnetic fields, destroying or altering the expected behavior of process in which is projected to be used. Quantum control could be used to correct that situation in several ways. Sometimes the user should be make some measurement upon the system to decide which is the best control scheme; other posibility is try to reconstruct the system using similar procedures without perturbate it. In the complete pictures both schemes are present. We will work first with pure systems studying advantages of different procedures. After, we will extend these operations when time of distortion is uncertain, generating a mixed state, which needs to be corrected by suposing the most probably time of distortion.



\end{abstract} 

\maketitle

\section{Introduction}

Physical elements of quantum information tasks are bounded to imperfections, decoherence or inclusively self distortion by natural interactions between their parts. Under this perspective, it is necessary introduce alternate elements to the quantum information process which take care to correct these deviations from the expected ideal behavior. Classical control is the science which treat with this kind of problem, obtaining some information by measuring, analizing and feedbacking the system in order to come back it to this ideal behavior.

Nevertheless,  the classical control was developed for systems which should be tolerant to noise or fault parts, upon supposition that these systems are not altered by measuring or at least altered in a quantifiable amount. Nevertheless quantum systems haven't this classical behavior because this alteration upon measurement is not completely quantifiable. In quantum mechanics, some kind of control is based on to exploit the properties of system in order to driving it \cite{mielnik1, fernandez1, delgado1, delgado2, delgado3}. But process to introduce control schemes in a classical sense on qubits has been recently developed taking measurements and giving feedback to them \cite{wang1, yuang1, branczyk1, xi1}.

In this paper we assume a double task. Mainly to extend the control process proposed by Bra\'nczyk \cite{branczyk1} and Xi \cite{xi1} to bipartite states (particularly in entangled states which are highly important in quantum computation and quantum information processes), but at the same time to study the effect of self interactions in noise or distortion introduction to ideal systems because of interaction between different parts of it.

The paper is organized as follows. Section II is devoted to explain Ising interaction model to use as immediate and temporal source of distortion after of his creation or preparation. Section III shows the behavior of initial states which motivate the necessity to introduce quantum control. In section IV and V, basic and simple control and his performance are discussed for pure states. Section VI extends some findings for mixed states and discusses the following program to improve the results.

\section{Ising model and evolution}

Ising model is motivated mainly by far-field strength of a dipole magnetic interaction between two particles. The classical Ising model is:

\begin{equation} E=-J \vec{\mathbf{s}}_1 \cdot \vec{\mathbf{s}}_2
\end{equation}

Nevertheless that this kind of interactions were first used in statistical physics to describe the magnetic behavior of lattices in different ways precisely by Ising \cite{ising1, brush1} and after by Heisenberg \cite{baxter1} in quantum mechanics, the interaction in the most symmetrical ways has been recently considered by other authors  to study transference and control of entanglement in bipartite qubits \cite{terzis1} and lattices \cite{stelma1, novotny1}. In addition, Cai \cite{cai1} has considered an more general model in order to study the relation between entanglement and local information.

Some interesting properties for this model in three dimensions on just two part systems and adding an inhomogeneous magnetic field in the $z$ direction has been recently studied \cite{delgado3}, obtaining a description of evolution and showing specific properties between entanglement and control. We will use this last model for interaction between particles:

\begin{equation} \label{hamiltonian}
H= -J \vec{\mathbf{\sigma}}_1 \cdot \vec{\mathbf{\sigma}}_2+B_1 {\sigma_1}_z +B_2 {\sigma_2}_z
\end{equation} 

\noindent which have the evolution operator \cite{delgado3} in Dirac notation:

\begin{eqnarray} \label{evolop}
U(t)&=&e^{-it'(b_+-j)} \left| 0_10_2 \right> \left< 0_10_2 \right| \nonumber \\ \nonumber
&&+ e^{-it'j}(\cos t'-i b_- \sin t') \left| 0_11_2 \right> \left< 0_11_2 \right| \\ \nonumber
&&+ i 2 j e^{-it'j}\sin t' \left| 0_11_2 \right> \left< 1_10_2 \right| \\ \nonumber
&&+ i 2 j e^{-it'j}\sin t' \left| 1_10_2 \right> \left< 0_11_2 \right| \\ \nonumber
&&+ e^{-it' j}(\cos t'+i b_- \sin t') \left| 1_10_2 \right> \left< 1_10_2 \right| \\ \nonumber
&&+e^{it'(b_++j)} \left| 1_11_2 \right> \left< 1_11_2 \right| \\ 
\end{eqnarray}

\noindent with: 

\begin{eqnarray} \label{rescaling}
& b_+=B_+/R, b_-=B_-/R \in [-1,1], \nonumber \\ 
& j=J/R \in [0,1/2], t'=Rt 
\end{eqnarray}.

\noindent where: $B_+=B_1+B_2$, $B_-=B_1-B_2$, $R=\sqrt{B_-^2+4J^2}$. We will drop the prime in the time in the following. 

\section{Initial states and distortion}

\subsection{General scheme of control}

The problem which we are dealing is depicted in figure 1. A system generates with the same probability one state at time from a set of two non-equivalent entangled states. We don't know which was the state produced. After of that, some kind of noise or distortion is introduced on the state produced, emerging to observer as a modified state which needs to be analyzed by measurement in order to reconstruct it into the original in the best possible way. We center our attention on the following alternative situations in presence of some uncontrolled inhomogeneus magnetic field, which induces distortion by some time $t$:

\begin{itemize}
\item {\bf Situation 1:} the observer tries to correct the distortion begining in some definite time still in presence of the original field, thinking that distortion has been acting by time $t$, or 
\item {\bf Situation 2:} after of distortion, suddenly, the two particles are set far away one of each other, stopping Ising and magnetic field interactions, then the observer begins the correction of the state
\end{itemize}

\subsection{Initial states}

If we begin with the following distinguishable bipartite states $\left| \beta_{1} \right>$  and  $\left| \beta_{2} \right>$:

\begin{eqnarray}
\left| \beta_{1} \right> &=& \frac{1}{\sqrt{2}} (\left| \phi_{1} \right> \left| \phi_{2} \right>+\left| \eta_{1} \right> \left| \eta_{2} \right>)  \nonumber \\ 
\left| \beta_{2} \right> &=& \frac{1}{\sqrt{2}} (\left| \varphi_{1} \right> \left| \varphi_{2} \right>+\left| \mu_{1} \right> \left| \mu_{2} \right>)
\end{eqnarray}

\noindent with:

\begin{eqnarray}
\left| \phi_i \right> &=& \cos \frac{\theta}{2} \left| 0_i \right>+\sin \frac{\theta}{2} \left| 1_i \right>, \nonumber \\ \nonumber
\left| \eta_i \right> &=& \sin \frac{\theta}{2} \left| 0_i \right>-\cos \frac{\theta}{2} \left| 1_i \right> \\ \nonumber
\left| \varphi_i \right> &=& \sin \frac{\theta}{2} \left| 0_i \right>+\cos \frac{\theta}{2} \left| 1_i \right>, \\ 
\left| \mu_i \right> &=& \cos \frac{\theta}{2} \left| 0_i \right>-\sin \frac{\theta}{2} \left| 1_i \right> 
\end{eqnarray}

\noindent for $i=1,2$, which is a set of orthonormal states. It's noticeable that:

\begin{eqnarray} \label{initialstates}
\left| \beta_1 \right> &=& \left| \beta_{00} \right> \nonumber \\ 
\left| \beta_2 \right> &=& \sin \theta \left| \beta_{01} \right> - \cos \theta \left| \beta_{11} \right> 
\end{eqnarray}

\noindent with $\left| \beta_{ij} \right>$ and $i,j \in \{0,1 \}$, the standard Bell states.

Nevertheless last formulae give some explanation for  constitution of the bipartite states in terms of certain pure entangled states, the most important technical aspect is that parameter $\theta$ lets us have a wider spectrum of states to compare. Trace distance has been used to measure distinguishability between two states \cite{fuchs1}. For our states, this measure is given by $\delta(\rho_1,\rho_2)=\frac{1}{2} \rm{Tr} |\rho_1-\rho_2|=\sin^2 \theta$, where $\rho_1=\left|\beta_1 \right> \left< \beta_1 \right|$ and $\rho_2=\left|\beta_2 \right> \left< \beta_2 \right|$. So, the last state, $\left| \beta_2 \right>$, goes from a very similar state to $\left| \beta_1 \right>$ when $\theta=0$ until a very different state when $\theta=\pi/2$.

\subsection{Distortion and evolution of entanglement}

In the scheme introduced by Bra\'nczyk \cite{branczyk1} and Xi \cite{xi1} the task is identify between two similar states after that some kind of noise or distortion is introduced. In an alternative way, this paper consider another kind of noise or distortion that those authors, because dephasing noise and bit flipping noise don't affect the entangled bipartite system which we are considering. 

Ising interaction leave invariant our initial states (\ref{initialstates}) in absence of magnetic field \cite{delgado3}, but the aim of this work is inquire about how the additional presence of the field (by example an parasite magnetic field) introduces an effective distortion. According with hamiltonian (\ref{hamiltonian}), the evolution of each state after of time $t$ of interaction we have:

\begin{eqnarray} 
\left| \beta_1' \right> &=&\frac{1}{\sqrt{2}} (e^{-it(b_+-j)} \left| 0_1 0_2 \right>+e^{it(b_++j)} \left| 1_1 1_2 \right>) \nonumber \\ \nonumber
\left| \beta_2' \right> &=&\frac{1}{\sqrt{2}} (-e^{-it(b_+-j)} \cos \theta \left| 0_1 0_2 \right> + \\ \nonumber
&& e^{-itj} \sin \theta (2ij \sin t+\cos t-ib_- \sin t)\left| 0_1 1_2 \right>+ \\ \nonumber
&& e^{-itj} \sin \theta (2ij \sin t+\cos t+ib_- \sin t)\left| 1_1 0_2 \right>+ \\ \nonumber
&& e^{it(b_++j)} \cos \theta \left| 1_1 1_2 \right>) \\
\end{eqnarray}

Calculating $\rho_1' \equiv \left| \beta_1' \right> \left< \beta_1' \right|$ and $\rho_2' \equiv \left| \beta_2' \right> \left< \beta_2' \right|$, we can verify that this distortion is trace distance preserving: $\delta(\rho_1',\rho_2')=\sin^2 \theta$. The first evolved state is maximally entangled always but the second not; the Schmidt coefficients for this last state become:

\begin{eqnarray}
\lambda_{1,2}&=&\frac{1}{2} \left( 1 \pm 
\sqrt{A+B \sin^2 2 \theta} \right)  \nonumber \\ \nonumber
{\rm with:} \\ \nonumber
A&=&16j^2(1-4j^2)\sin^4 t \sin^4 \theta \\ \nonumber
B&=&  \sin^2 2jt + 4j^2 \sin^2 t \cos 4jt - \\ 
&&j \sin 2jt \sin 4jt  
\end{eqnarray}

\noindent which are periodic just if $j \in \mathbb{Q}$ (see also \cite{delgado3}), so the argument of square root is equal to zero (maximally entangled) only in non periodic times and distorted states will become normally partial entangled in addition \cite{delgado3}. 

\section{Basic control of Ising distortion for pure states}

A very basic control as was introduced in \cite{delgado3} could solve the distortion introduced in before section. Here two aspect were introduced: the first one is the periodical behavior of evolution operator (\ref{evolop}) for $j \in \mathbb Q$ and the second is the concept of evolution loops (or more precisely, the operation of time reversion introduced in \cite{mielnik1, delgado1}, which completes a definite evolution into an evolution loop).

\subsection{Situation 1}
For the situation 1 depicted in section III, observer still could use the inhomogeneus field to induce a reconstruction of the original state. This scheme of reconstruction is independent of the initial state. So, observer could  retrieve the original state but without know it. The scheme is simple, by adding an extra homogeneous magnetic field by some time $T$ (note that this traduces just in a greater average field $b_+ \rightarrow b_+ + \delta b_+$ because $b_-$ remains unchanged and so the renormalization (\ref{rescaling}) remains valid) we obtain the evolution:

\begin{equation} \label{evolsit1}
U(t+T)=U_{b_+ + \delta b_+}(T)U_{b_+}(t)
\end{equation}

\noindent after of time $t+T$. By select the parameters:

\begin{eqnarray} \label{solsit1}
T &=& n \pi-t \nonumber \\
\delta b_+ &=& \frac{\pi(2m-n(b_+ - 2j+1))}{T} \nonumber \\
\frac{s}{2n} &=& Q(j) \nonumber \\
{\rm with:} && n, m, s \in \mathbb Z
\end{eqnarray}

\noindent where $Q(j)$ means some rational approximation to $j$ in terms of $n$ as denominator, by the selection of $n$ and $s$, so closest to $j$ as be possible (in case that $j \in \mathbb Q$, then $j=Q(j)$ and selection is direct but with several options), we obtain (until unitary factors) a quasi evolution loop: $U(t+T)=I'$ with $I'$ as diagonal matrix: $I'={\rm diag}(1,1,1,e^{4 i n \pi \delta})$, where $j=Q(j)+\delta$. 

Note that $I'$ expression doesn't converge to $I$ always because precision of rational approximation to $j$ it's inverse to $n$, so $n \delta$ doesn't converge to zero (at least that $j$ will be rational). Otherwise, if it's desirable reprepare the state as soon as possible, then $n \pi$ should be slightly bigger than $t$, limiting the precision of $Q(j)$. In addition, $\delta b_+$ will be as stronger as $2m-n(b_+ - 2j+1) \approx m+s-n(b_+ +1) $ needs. This it's a possible way to reprepare the states in situation 1, indepently of their initial state itself.

\subsection{Situation 2}
Situation 2 is more complicated because far away of magnetic field which causes the distortion, we can't take advantage of properties of magnetic field which generates distortion. Otherwise, it's improbable that observer could reproduce the original conditions in the two places that particles have been located now, so for this reason we can make just local operations to try to return the state to the original (it means the use of (\ref{hamiltonian}) with $J=0$). 

Applying magnetic fields to each particle $B_1'$ and $B_2'$ for a time $T$, and taking $B_+'=B_1'+B_2'$, $B_-'=B_1'-B_2'$:

\begin{equation} \label{evolsit2}
U(t+T)=U_{B_+',B_-',J=0}(T)U_{B_+,B_-,J}(t)
\end{equation}

\noindent  we found that unfortunately there are not a single selection of $B_+', B_-'$ to construct a time reversion operation, neither to reconstruct both states with exactly the same requirements (so probably, the observer could to require make a measurement to try to know what state should reconstruct and select the best conditions).

For the best of scenarios we obtain as reprepared closest states to $\left| \beta_1 \right>$ and $\left| \beta_2 \right>$ (dropping some unitary factors):

\begin{eqnarray} \label{reconstructedstates}
\left| \beta_1'' \right>&=&\left| \beta_1 \right> \nonumber \\
\left| \beta_2'' \right>&=&e^{i \Delta} \frac{\sin \theta}{\sqrt{2}}(r \left| 0_1 1_2 \right> + r' \left| 1_1 0_2 \right> )-\cos \theta \left| \beta_{11} \right> \nonumber \\
\end{eqnarray} 

\noindent where:

\begin{eqnarray} \label{parameterssit1}
r&=&\sqrt{1-\frac{4JB_-}{R^2}\sin^2 R t} \nonumber \\
r'&=&\sqrt{1+\frac{4JB_-}{R^2}\sin^2 R t} \nonumber \\
\Delta&=&-(m \pi +\frac{1}{2}f(B_-t,2Jt) ) \nonumber \\
f(B_-t,2Jt)&=&\phi-\phi'-4 J t \nonumber \\
\phi&=&\arctan (\frac{B_--2J}{R} \tan Rt) \nonumber \\
\phi'&=&\arctan (\frac{B_-+2J}{R} \tan Rt) \nonumber \\
\end{eqnarray}

One common condition to reconstruct both states (and sufficient for $\left| \beta_1 \right>$) as in (\ref{reconstructedstates}), it's just:

\begin{eqnarray} \label{solsit2a}
B_+'T &=&n \pi-B_+ t
\end{eqnarray}

\noindent with $n \in \mathbb Z$. Now, to reconstruct approximately  $\left| \beta_2 \right>$ as in (\ref{reconstructedstates}), observer requires in addition:

\begin{eqnarray} \label{solsit2b}
B_-' T &=& (m+n)\pi-\frac{1}{2}(\phi+\phi') \nonumber \\
\end{eqnarray}

\noindent with $m \in \mathbb Z$. It's covenient that:

\begin{eqnarray} \label{cond1}
f(B_-t,2Jt)=2m\pi
\end{eqnarray}

\noindent nevertheless not always it's possible (Fig. 2). Additional attempts to reconstruct completely $\left| \beta_2 \right>$ require to have control and freedom to select $t$, but this isn't. In general, while sooner the process of reconstruction, the magnetic field needed for repreparation should be stronger and more inhomogeneous because of (\ref{solsit2a}) and (\ref{solsit2b}). Any way the conclusion is that fulfilling the conditions for reconstruct the second state we automatically fulfill the condition for the first state, so the observer doesn't require to take an intermediate measure.

Particularly, when $Rt=p \pi$ with $p \in \mathbb Z$, then $r=r'=1$ giving a perfect reconstruction of the state (note that altenatively if field is homogeneous, $r=r'=1$. But in addition is needed that:

\begin{eqnarray}
\frac{2J}{R}=\frac{1}{\sqrt{1+(B_-/2J)^2}}=\frac{m-m'}{p} 
\end{eqnarray}
 
\noindent  which despite that can be fulfilled with sufficiently precision by a rational approximation: $Q(\frac{2J}{R})=\frac{m-m'}{p}$, is useless because the term $2Jt=\frac{2J}{R} p \pi$ in the expressions of state evolution can't be delimited because $Q(\frac{2J}{R})$ is inversely proportional to $p \pi$ (just if $\frac{2J}{R} \in \mathbb Q$ the last process become useful). 

\section{Simple control for pure states: discriminate and reprepare}

\subsection{Program of control}

Now, instead of the last for situations 1 and 2, we will assume that observer is interested in to make a program of control with a measurement intermediate in order to know what state was originally created and reconstruct it in the most accurate or convenient way. Following to Bra\'nczyk \cite{branczyk1} and Xi \cite{xi1}, in this stage, some measurement should be made on each qubit in order to obtain some information about the actual state and so, to identify with certain probability which was the original created by the system. In a general procedure, we need to define a POVM as:

\begin{equation}
\left\{ \left| \delta_1 \delta_2 \right>, \left| \epsilon_1 \epsilon_2 \right>, \left| \delta_1 \epsilon_2 \right>, \left| \epsilon_1 \delta_2 \right> \right\}
\end{equation} 

\noindent where:   $\left\{ \left| \delta_1 \right>,\left| \epsilon_1 \right> \right\}$ and   $\left\{ \left| \delta_2 \right>,\left| \epsilon_2 \right> \right\}$ are orthonormal states for each part. The Helmstrom probabilities (success probabilities) are:

\begin{eqnarray} \label{helmstrom}
{P_H}_1=\left| \left< \beta_1' | \delta_1 \delta_2 \right> \right|^2+
\left| \left< \beta_1' | \epsilon_1 \epsilon_2 \right> \right|^2 \\ \nonumber
{P_H}_2=\left| \left< \beta_2' | \delta_1 \epsilon_2 \right> \right|^2+
\left| \left< \beta_2' | \epsilon_1 \delta_2 \right> \right|^2
\end{eqnarray}

\noindent where  $\left| \beta_1' \right>$, $\left| \beta_2' \right>$ are the states after of distortion arising from the original states $\left| \beta_1 \right>$, $\left| \beta_2 \right>$  trough of the Hamiltonian of interaction. With this, the average fidelity becomes:

\begin{eqnarray} \label{fidelity}
\overline{F}&=&\frac{1}{2} 
\left( 
\left< \beta_1 \left| (  \left| \beta_1'' \left> \left< \beta_1'' \left| {P_H}_1 + \right| \beta_2'' \right> \right<  \beta_2'' \right| (1-{P_H}_1) ) \right| \beta_1 \right> 
\right)+ \nonumber \\
&& \frac{1}{2} 
\left( 
\left< \beta_2 \left| (  \left| \beta_1'' \left> \left< \beta_1'' \left| (1-{P_H}_2) + \right| \beta_2'' \right> \right<  \beta_2'' \right| {P_H}_2) \right| \beta_2 \right> 
\right) \nonumber \\
&=& \frac{1}{2} \left( 
P_{H_1} \left| \left< \beta_1 | \beta_1'' \right> \right|^2+
(1-P_{H_1}) \left| \left< \beta_1 | \beta_2'' \right> \right|^2 \right) + \nonumber \\
& & \frac{1}{2} \left( P_{H_2} \left| \left< \beta_2 | \beta_2'' \right> \right|^2+
(1-P_{H_2}) \left| \left< \beta_2 | \beta_1'' \right> \right|^2
\right)
\end{eqnarray}

\noindent with $\left| \beta_1'' \right>$, $\left| \beta_2'' \right>$ the states to reprepare after of measurement (by example following the procedures delined in before section). Note that in this scheme we don't work with mixed states as in \cite{branczyk1, xi1}. So, our task is to find which are the best states:

\begin{equation}
\left| \delta_1 \right>, \left| \epsilon_1 \right>, \left| \delta_2 \right>, \left| \epsilon_2 \right>, \left| \beta_1'' \right>, \left| \beta_2'' \right>
\end{equation}

\noindent to measure and reprepare the system in order to maximize the average fidelity.

\subsection{Intuitive approaches}

As an example, in a very intuitive approach we take: 

\begin{eqnarray}
\left| \delta_1 \right>=\left| 0_1 \right>, \left| \epsilon_1 \right>=\left| 1_1 \right>,
\nonumber \\
\left| \delta_2 \right>=\left| 0_2 \right>, 
\left| \epsilon_2 \right>=\left| 1_2 \right>,
\nonumber \\
\left| \beta_1'' \right>=\left| \beta_1 \right>, \left| \beta_2'' \right>=\left| \beta_2 \right>
\end{eqnarray}

\noindent last reprepared states cold be achieved with procedures delined in subsection A. How in our case  $\left< \beta_1 | \beta_2 \right>=0$, some calculations show that:

\begin{eqnarray} \label{fdr1}
{P_H}_1&=&\left| \left< \beta_1' | 0_1 0_2 \right> \right|^2+
\left| \left< \beta_1' | 1_1 1_2 \right> \right|^2=1 \\ \nonumber
{P_H}_2&=&\left| \left< \beta_2' | 0_1 1_2 \right> \right|^2+
\left| \left< \beta_2' | 1_1 0_2 \right> \right|^2= \sin^2 \theta \\ \nonumber
&& \Rightarrow \overline{F} =  F_{DR_1}=\frac{1}{2} (1+\sin^2 \theta)
\end{eqnarray}

For comparison, we take the Do-nothing scheme (as is used in [1, 2], this is a scheme where we don't make re-preparation and Helmstrom probabilities are 1) and we obtain:

\begin{eqnarray} \label{fn}
\overline{F} & = & F_N =\frac{1}{2} \left| \left< \beta_1 | \beta_1' \right> \right|^2+ \frac{1}{2} \left| \left< \beta_2 | \beta_2' \right> \right|^2 \nonumber \\
&=& \frac{1}{2} ( \cos^2 b_+ t ( 1+ \cos^4 \theta) + \nonumber \\
& & \frac{1}{2} \cos b_+ t ( \cos t \cos 2 j t + 2 j \sin t \sin 2 j t)\sin^2 2 \theta + \nonumber \\
& &  ( 4 j^2 \sin^2 t + \cos^2 t )\sin^4 \theta )\nonumber \\
\end{eqnarray}

This fidelity doesn't depend on the inhomogeneity of field but its strength $b_+$. Fidelity for both procedures is shown in figure 3, as a function of parameters $\theta$, $b_+$. We note that first procedure is quit inefficient when $\theta \approx 0$ (with initial states very similar) for different values for duration of interaction.

\subsection{Suboptimal control}

In general, our optimization problem is to take:

\begin{eqnarray} \label{measure}
\left| \delta_1 \right> &=& \cos \frac{\theta_1}{2} \left| 0_1 \right>+ e^{i \alpha_1} \sin \frac{\theta_1}{2} \left| 1_1 \right>  \nonumber \\
\left| \epsilon_1 \right> &=& \sin \frac{\theta_1}{2} \left| 0_1 \right>- e^{-i \alpha_1} \cos \frac{\theta_1}{2} \left| 1_1 \right> \nonumber \\
\left| \delta_2 \right> &=& \cos \frac{\theta_2}{2} \left| 0_2 \right>+ e^{i \alpha_2} \sin \frac{\theta_2}{2} \left| 1_2 \right>  \nonumber \\
\left| \epsilon_2 \right> &=& \sin \frac{\theta_2}{2} \left| 0_2 \right>- e^{-i \alpha_2} \cos \frac{\theta_2}{2} \left| 1_2 \right> \nonumber \\
\end{eqnarray}

\noindent and to find the best parameters $\theta_1, \theta_2, \alpha_1, \alpha_2$ with an adequate selection of $\left| \beta_1'' \right>, \left| \beta_2'' \right>$ to maximize $\overline{F}$ in (\ref{fidelity}). 

The last just in case that we agree take the most coventional way of mesurement, it means, take a measure on each part. Other ways of measurment can be implemented taking measurements which involve both parts at the time. 

In the case that we will reprepare the measured state upon the result with the original states (\cite{branczyk1} and \cite{xi1} shown that this is not the optimal way to maximize $\overline{F}$) the optimization problem becomes:

\begin{eqnarray} \label{fdr2a}
F_{DR_2} & = &  {\rm Max} \{ \overline{F} \} \nonumber \\
{\rm with:} & &  \nonumber \\
\overline{F} & = & \frac{1}{2} \left( 
P_{H_1} \left| \left< \beta_1 | \beta_1' \right> \right|^2+
(1-P_{H_1}) \left| \left< \beta_1 | \beta_2' \right> \right|^2 \right) + \nonumber \\
& & \frac{1}{2} \left( P_{H_2} \left| \left< \beta_2 | \beta_2' \right> \right|^2+
(1-P_{H_2}) \left| \left< \beta_2 | \beta_1' \right> \right|^2 \right) \nonumber \\
\end{eqnarray}

\noindent with $P_{H_1}, P_{H_2}$ as in (\ref{helmstrom}) and $\theta_1, \theta_2, \alpha_1, \alpha_2$ as in (\ref{measure}). This optimization problem is in general difficult of solve because of large number of parameters. Nevertheless for the case without magnetic field, it reduces to:

\begin{eqnarray}
& {\rm Max} \{ \cos \alpha_1 \sin \theta_1 (\cos \theta_2 \sin 2 \theta + 
2 \cos \alpha_2 \cos^2 \theta \sin \theta_2) +   \nonumber \\
& \cos \theta_1 (2 \cos \theta_2 \sin^2 \theta + 
      \cos \alpha_2 \sin 2 \theta \sin \theta_2) \}  \nonumber \\
\end{eqnarray}

\noindent taking derivatives with respect to $\theta_1, \theta_2, \alpha_1, \alpha_2$ and solving the optimization problem, it conduces to 108 solutions giving the seven critical fidelities: $0 \le \frac{1}{2}(1-\sin\theta)  \le \frac{1}{2}(1-\sin^2\theta)  \le \frac{1}{2}  \le \frac{1}{2}(1+\sin^2 \theta)  \le \frac{1}{2}(1+\sin\theta)  \le 1$. From these solutions, 8 solutions emerge with $\overline{F}=1$, which generate the two physically different set of measurements shown in Table I, named A and B in the following.

\begin{table} \label{tabla1}
\centering
\begin{tabular}[b]{|c|c|}
\hline
{\bf Type} & {\bf POVM} \\
\hline
  & $\left| \delta_1 \right> = \cos \frac{1}{4} (\pi-2 \theta) \left| 0_1 \right> + \sin \frac{1}{4} (\pi-2 \theta) \left| 1_1 \right> $  \\
A & $\left| \epsilon_1 \right> = \sin \frac{1}{4} (\pi-2 \theta) \left| 0_1 \right> - \cos \frac{1}{4} (\pi-2 \theta) \left| 1_1 \right> $  \\
  & $\left| \delta_2 \right> = \cos \frac{1}{4} (\pi-2 \theta) \left| 0_2 \right> + \sin \frac{1}{4} (\pi-2 \theta) \left| 1_2 \right> $  \\
  & $\left| \epsilon_2 \right> = \sin \frac{1}{4} (\pi-2 \theta) \left| 0_2 \right> - \cos \frac{1}{4} (\pi-2 \theta) \left| 1_2 \right> $  \\
\hline
  & $\left| \delta_1 \right> = \cos \frac{1}{4} (\pi+2 \theta) \left| 0_1 \right> - \sin \frac{1}{4} (\pi+2 \theta) \left| 1_1 \right> $  \\
B & $\left| \epsilon_1 \right> = \sin \frac{1}{4} (\pi+2 \theta) \left| 0_1 \right> + \cos \frac{1}{4} (\pi+2 \theta) \left| 1_1 \right> $  \\
  & $\left| \delta_2 \right> = \cos \frac{1}{4} (\pi+2 \theta) \left| 0_2 \right> - \sin \frac{1}{4} (\pi+2 \theta) \left| 1_2 \right> $  \\
  & $\left| \epsilon_2 \right> = \sin \frac{1}{4} (\pi+2 \theta) \left| 0_2 \right> + \cos \frac{1}{4} (\pi+2 \theta) \left| 1_2 \right> $  \\
\hline
\end{tabular}
\caption{The two solutions for POVM (\ref{measure}) which maximize $\overline{F}=1$ in absence of magnetic field.}
\end{table}

Of course, in this scheme states are not distorted after of Ising interaction because their invariance \cite{delgado3}. In addition, this scheme is limited to the absence of magnetic field, nevertheless it will be interesting analyzing how this set of measurements works by extension when magnetic field is present. Applying POVM type A or B of Table I in the presence of an inhomogeneous magnetic field and repreparating the system with the original states we obtain:

\begin{eqnarray}
F_{A,B} & = &
\frac{1}{8} (4 \cos^2 b_+ t (1+\cos^4 \theta) + \nonumber \\
& & ((1+4j^2)+ 8 \cos b_+ t \cdot \nonumber \\
& & ((1+2j) \cos(1-2j)t+(1-2j) \cos(1+2j)t) + \nonumber \\
& & + 4 \sin^2 b_+ t -2(1-4j^2)) \cos^2 \sin^2 + \nonumber \\
& & (3(1+4j^2)+4 \sin^2 b_+ t) \sin^4 \theta)
\end{eqnarray}

Figure 3 shows the comparison between $F_{DR_1}, F_N, F_{A,B}$  for an illustrative time of interaction and $j$ value. $F_{A,B}$ reduces when $\theta \rightarrow 0$ to the same value of $F_N$:

\begin{eqnarray} \label{lim1}
\lim_{\theta \rightarrow 0} F_{A,B}=\lim_{\theta \rightarrow 0} F_N=\cos ^2 b_+ t
\end{eqnarray}

\noindent which shows the oscillatory behavior of both procedures respect magnetic field strenght. Similarly, when $\theta \rightarrow \frac{\pi}{2}$:

\begin{eqnarray} \label{lim2}
\lim_{\theta \rightarrow \frac{\pi}{2}} F_{A,B}=\lim_{\theta \rightarrow \frac{\pi}{2}} F_{DR_1}=1
\end{eqnarray}

Finally we can observe that:

\begin{eqnarray} \label{lim3}
\lim_{\theta \rightarrow \frac{\pi}{2}} F_N=\frac{1}{2}(1+(4j^2-1) \sin^2 t+\cos^2 b_+ t)
\end{eqnarray}

In this sense, both procedures $F_{DR_1}$ and $F_{A,B}$ are covering partially the control with fidelity at least of $\frac{1}{2}$. For reference, we define in the following:

\begin{eqnarray} \label{fso}
F_{SO}={\rm Max}\{F_{DR_1},F_{A,B} \} \\ \nonumber
\end{eqnarray}

\subsection{Optimal control}
As was said, obtain $F_{DR_2}$ analytically is actually difficult. Nevertheless this problem could be solved numerically. Taking a region $\Re \equiv [\theta_{min},\theta_{max}] \times [b_{+_{min}},b_{+_{max}}]$ for some $t$ and $j$, we can to seek the best values for $\theta_1, \theta_2, \alpha_1, \alpha_2$ for each $(\theta,b_+)$ with some appropriate numerical algorithm.

For the example depicted before, we obtain $F_{DR_2}$ compared with $F_{SO}$ shown in figure 3. Observe now the increased fidelity in around of 50\% around of $b_+=\frac{(2n+1)\pi}{2t}, n \in \mathbb Z$ and slightly by 15\% near of $b_+=\frac{n \pi}{2t}, n \in \mathbb Z$ and $\theta=\frac{\pi}{4}$. With this, around of 80\% of this region have a fidelity up of 0.8 (Fig. 4).

\section{Control over mixed states created by stochastic Ising distortion}

\subsection{Stochastic Ising interaction and entanglement evolution}

In the last section was considered the case when distortion is applied with certainty after of the process for produce the states. Nevertheless, if we have some stochastic component about this distortion, additional considerations appear. Instead to follow Bra\'nczyk \cite{branczyk1} and Xi \cite{xi1} by considering the application of distortion with some probability (which is more situable for dephasing noise or bit flipping noise), we will think in some normal distribution for the duration time of Ising interaction around of $t_0$ with a dispersion $s$:

\begin{eqnarray} \label{normal}
f(t)=\frac{1}{\sqrt{2 \pi}s}e^{-\frac{(t-t_0)^2}{2 s^2}}
\end{eqnarray}

With this, the states after of interaction are:

\begin{eqnarray} \label{mixedstate}
\rho_i'=\int_{-\infty}^{\infty} f(t) U(t) \rho_i U ^\dagger(t) dt
\end{eqnarray}

\noindent for $i=1,2$, with: $\rho_i=\left| \beta_i \right> \left< \beta_i \right|$ and $U(t)$ given by (\ref{initialstates}) and (\ref{evolop}) respectively. We assume that $t_0/s \gg 1$ in order to take these integral limits. Expressions for this last states are so large to include here (specially $\rho_2'$), nevertheless they are analytically achieved. It's well known that operators:

\begin{eqnarray} \label{entwitness}
W_{ij}& \equiv & 1-2 \left| \beta_{ij} \right> \left< \beta_{ij} \right| \nonumber \\
{\rm with:} & & i,j \in {0,1}
\end{eqnarray}

\noindent are entanglement witnesses \cite{horo1}. Applying this operators to $\rho_1'$ we obtain that ${\rm Tr}(W_{ij}\rho_1')=1>0$, except for:

\begin{eqnarray}
{\rm Tr}(W_{00}\rho_1')&=&- e^{-2b_+^2s^2}\cos2b_+t_0 \nonumber \\
{\rm Tr}(W_{10}\rho_1')&=&e^{-2b_+^2s^2}\cos2b_+t_0 \nonumber \\
\end{eqnarray}

\noindent expliciting that $\rho_1$ is an entangled state independently of values for $b_+$ and $t_0$. Note that they go to zero when $s \rightarrow \infty$. Similarly for $\rho_2'$:

\begin{eqnarray}
{\rm Tr}(W_{00}\rho_2')&=&1-\cos^2 \theta (1-e^{-2b_+^2s^2}\cos2b_+t_0) \nonumber \\
{\rm Tr}(W_{01}\rho_2')&=&1-\sin^2 \theta ((1+4j^2)+ \nonumber \\
& & (1-4j^2)e^{-2s^2}\cos2t_0) \nonumber \\
{\rm Tr}(W_{10}\rho_2')&=&1-\cos^2 \theta (1+e^{-2b_+^2s^2}\cos2b_+t_0) \nonumber \\
{\rm Tr}(W_{11}\rho_2')&=&1-(1-4j^2)\sin^2 \theta (1-e^{-2s^2}\cos2t_0) \nonumber \\
\end{eqnarray}

\noindent which for some selection of parameters $b_+, s, j, \theta$ and $t_0$ become all positive, it means that $\rho_2$ is a separable states in such case. Particularly for $s \gg |b_+|^{-1}$ we will have entangled states if $\theta >\arcsin(1+4j^2)^{\frac{1}{2}}$.

\subsection{Results for fidelity in the classical procedures}

Now, Helmstrom probabilities are calculated as:

\begin{eqnarray} \label{helmstrommix}
{P_H}_1=\left| \left< \delta_1 \delta_2 |\rho_1' | \delta_1 \delta_2 \right> \right|^2+
\left| \left< \epsilon_1 \epsilon_2 |\rho_1' | \epsilon_1 \epsilon_2 \right> \right|^2 \\ \nonumber
{P_H}_2=\left| \left< \delta_1 \epsilon_2 | \rho_2' | \delta_1 \epsilon_2 \right> \right|^2+
\left| \left< \epsilon_1 \delta_2 | \rho_2'| \epsilon_1 \delta_2 \right> \right|^2
\end{eqnarray}

\noindent and the fidelity is:

\begin{eqnarray} \label{fidelitymix}
\overline{F}&=&
\frac{1}{2}  \left(
P_{H_1} {\rm Tr} ( \rho_1 \rho_1'') +
(1-P_{H_1}) {\rm Tr} (\rho_1 \rho_2'') \right) + \nonumber \\
& & \frac{1}{2} \left( P_{H_2} {\rm Tr} (\rho_2 \rho_2'') +
(1-P_{H_2}) {\rm Tr} (\rho_2 \rho_1'')
\right)
\end{eqnarray}

\noindent where $\rho_1'', \rho_2''$ are the reprepared states after of mesurement. Reapeating calculations of the before section we obtain $F_{DR_1}$ as in (\ref{fdr1}) because this fidelity was independent of $t$. This is noticeable because for $\theta \ge \frac{3\pi}{8}$ this is a very good fidelity and is independent of $s$. Nevertheless, an unknown hypotesis has been used here, the reprepared states are exactly the same that the original, which not always is possible, at least with the procedures developed in section IV. We will study this in following subsections.

For $F_N$, after of some calculations:

\begin{eqnarray} \label{fnmix}
F_N &=&
\frac{1}{4} (
(1+e^{-2b_+^2s^2} \cos 2b_+ t_0)(1+\cos^4 \theta)+ \nonumber \\
& & G_N(b_+,j,t_0,s) \cos^2 \theta \sin^2 \theta + \nonumber \\
& & ((1+4j^2)+(1-4j^2)e^{-2s^2} \cos 2t_0) \sin^4 \theta) \nonumber \\
\end{eqnarray}

\noindent with: 

\begin{eqnarray}
\lefteqn {G_N(b_+,j,t_0,s)=} \nonumber \\
& & (1-2j)(e^{-\frac{1}{2}(1+b_++2j)^2s^2} \cos(1+b_++2j)t_0 \nonumber \\
& & + e^{-\frac{1}{2}(1-b_++2j)^2s^2} \cos(1-b_++2j)t_0)+ \nonumber \\
& & (1+2j)(e^{-\frac{1}{2}(1+b_+-2j)^2s^2} \cos(1+b_+-2j)t_0 \nonumber \\
& & + e^{-\frac{1}{2}(1-b_+-2j)^2s^2} \cos(1-b_+-2j)t_0) \nonumber \\
\end{eqnarray}

This expression reduces to (\ref{fn}) when $s \rightarrow 0$ and to $F_N=\frac{1}{4}(1+\cos^4 \theta+(1+4j^2) \sin^4 \theta)\le \frac{3}{4}$ when $s \rightarrow \infty$. Last value is achieved when $j=\frac{1}{2},\theta=\frac{\pi}{2}$ (most distinguishable initial states and strong Ising interaction respect to $b_+$). 

Similar numerical procedures that those of before section could give us optimal control, but they depend on a lot parameters including $s$. We will skip this study, instead, turn our attention in more basic but still effective control procedures.

\subsection{Repreparating mixed states}

Some aspects should be said related with restrictions of repreparation for the mixed states. Before, formula for average fidelity should be simplified in case that observer decides not take an intermediate measurement to make decisions about control scheme (similarly as in the Do-nothing scheme). In such case, it results from (\ref{fidelitymix}):

\begin{eqnarray} \label{fidelitymixnomeasure}
\overline{F}&=&
\frac{1}{2}  \left(
{\rm Tr} ( \rho_1 \rho_1'') + {\rm Tr} (\rho_2 \rho_2'')
\right)
\end{eqnarray}

Clearly more complications or diminishing of fidelity arise when $P_{H_1} \ne 1, P_{H_2} \ne 1$, if repreparation is near of perfection. So, in our last study will suppose for simplicity that the observer does not need make a measure to adjust some parameters of repreparation (as was in procedures of section IV), which is perfectly made for some $t$ but working over a mixed state given for uncertainty of this parameter, as was depicted in the before subsections.

\subsubsection{Situation 1}
In the situation 1 depicted in section III, repreparation begins at some time completely defined by the observer, still under the action of the original magnetic field but with partial ignorance of the begining of influence of it. For this reason, how the observer don't have perfect control of time $t_0$, although he has planned make a repreparation in terms of (\ref{solsit1}) for time $t=t_0$ not always will correct perfectly the state to the original because this uncertainty. So:

\begin{eqnarray} \label{mixedstate1}
\rho_i''=U_{b_+ + \delta b_+}(T_{t_0})\rho_i'U^\dagger_{b_+ + \delta b_+}(T_{t_0})
\end{eqnarray}

\noindent with with $\rho_i'$ calculated as in (\ref{mixedstate}), $U_{b_+ + \delta b_+}(T_{t_0})$ given by (\ref{evolsit1}) and (\ref{solsit1}), for all parameters of repreparation control, ($T, \delta b_+$), calculated for $t=t_0$. Note that measurement it isn't necessary here because observer doesn't need to know the initial state to correct it. (\ref{fidelitymixnomeasure}) is in this case:

\begin{eqnarray} \label{fmixsit1}
F_1 &=&
\frac{1}{4} (
(1+e^{-2b_+^2s^2} \cos 4jn \pi)(1+\cos^4 \theta)+ \nonumber \\
& & G_1(b_+,j,t_0,s) \cos^2 \theta \sin^2 \theta + \nonumber \\
& & ((1+4j^2)+(1-4j^2)e^{-2s^2}) \sin^4 \theta) \nonumber \\
\end{eqnarray}

\noindent with: 

\begin{eqnarray}
\lefteqn{ G_1(b_+,j,t_0,s) =} \nonumber \\
& & (1-2j)e^{-\frac{1}{2}(1+b_++2j)s^2} \cos 4jn \pi + \nonumber \\ 
& & (1-2j)e^{-\frac{1}{2}(1-b_++2j)s^2} + \nonumber \\
& & (1+2j)e^{-\frac{1}{2}(1-b_+-2j)s^2} \cos 4jn \pi + \nonumber \\  
& & (1+2j)e^{-\frac{1}{2}(1+b_+-2j)s^2}  \nonumber \\  
\end{eqnarray}

Note that dependence on $t_0$ is just trough $n$ from (\ref{solsit1}). This scheme of repreparation it's equivalent to the Do-nothing scheme  when $t_0=n \pi$ and $(1+b_+-2j)t_0=2 p \pi$ with $p \in \mathbb Z$, then $F_1$ matches with $F_N$. Some important properties are:

\begin{eqnarray}
\lim_{s \rightarrow 0} F_1 & = & \lim_{s \rightarrow 0} F_N \le 1 \\
\lim_{s \rightarrow \infty} F_1 & \le & \frac{3}{4} \label{lim1infty}
\end{eqnarray}

\noindent where the meaning of the inequalities is that with adequate selection of parameters this values are reached upmost.

Figure 5 shows $F_N$ (black) togheter $F_1$ (gray) for $b_+=1, j=\frac{1}{6}$ as an example, for different values of $t_0=\frac{\pi}{2}, \frac{3\pi}{4}, \frac{7\pi}{4}$ depending of $\theta$ and $s$ (here, $n, m$ were selected to give the positive values of $T, b_+$ closest to zero). We take $s \in [0,\frac{t_0}{3}]$ as the maximum value for which  (\ref{normal}) remains valid. Note that in general by increasing $t_0$ the fidelity goes down as a result of periodicity of phenomena reported in \cite{delgado3}. In addition, the imprecision of repreparation in $F_1$ sometimes gives worst values that $F_N$. 

\subsubsection{Situation 2}
In the situation 2, after of interaction, distorted state remains unaltered in principle and observer could decide then to take a measurement to fit the best parameters for repreparation (this is not strictly necessary if he decides to apply magnetic fields as it's required by the second state as in (\ref{solsit2a}) and (\ref{solsit2b}), because authomatically condition for first state is fulfill). He has still some uncertainty about the real time $t_0$ of interaction. In this case:

\begin{eqnarray} \label{mixedstate2}
\rho_i''=U_{B_+',B_-',J=0}(T_{t_0}) \rho_i' U^\dagger _{B_+',B_-',J=0}(T_{t_0})
\end{eqnarray}

\noindent with $U(T)$ given by (\ref{evolop}) but with $J=0$ as was depicted in the section IV.B. (\ref{fidelitymixnomeasure}) becomes for this case:

\begin{eqnarray} \label{fmixsit2}
F_2 &=&
\frac{1}{4} (
(1+e^{-2b_+^2s^2})(1+\cos^4 \theta)+ \nonumber \\
& & G_2(b_+,b_-,j,t_0,s) \cos^2 \theta \sin^2 \theta + \nonumber \\
& & ((1+4j^2 \cos \Delta)+ b_- e^{-2s^2} \sin \Delta \sin 2t_0 +\nonumber \\
& & (1-4j^2)e^{-2s^2}\cos \Delta \cos 2t_0 ) \sin^4 \theta) \nonumber \\
\end{eqnarray}

\noindent with: 

\begin{eqnarray}
\lefteqn{ G_2(b_+,b_-,j,t_0,s) =} \nonumber \\
& & \sum_{p,q,r \in \{-1,1\}} (-1)^m(1-qrb_--2qj) \cdot \nonumber \\ 
& & e^{-\frac{1}{2}(1+pb_+2qj)s^2} \cos(q+2j+r\frac{\Delta}{2}) \nonumber \\
\lefteqn{\Delta=\phi+\phi'}
\end{eqnarray}

\noindent where $m, \phi$ and $\phi'$ are the parameters in (\ref{parameterssit1}) with $t=t_0$. It's noticeable that repreparation in terms of subsection IV.B it's equivalent to Do-nothing scheme when $b_+ t_0=2 s \pi$ and $\Delta=2 m \pi$, in that situation (\ref{fnmix}) and (\ref{fmixsit1}) match. Some important properties are:

\begin{eqnarray}
\lim_{s \rightarrow 0} F_2 & \le & 1 \\
\lim_{s \rightarrow \infty} F_2 & \le & \frac{3}{4} \label{lim2infty}
\end{eqnarray}

\noindent where the meaning of the inequalities is that with adequate selection of parameters this values are reached upmost.

Figure 5 shows complete comparison with the three last control schemes, exhibiting an aparent superiority of $F_2$ because its better repreparation. This is true at least for smaller values of $t_0$, nevertheless for intermediate values there are a combination for superiority in patches. It's clear the decreasing fidelity for $s \rightarrow \infty$ as is expected.

\subsection{Final remarks}
Comparing our results for $F_N, F_1$ and $F_2$ we note that all have the form:

\begin{eqnarray} \label{fmixsit2}
F &=&
\frac{1}{4} (
A(1+\cos^4 \theta)+ \nonumber \\
& & G \cos^2 \theta \sin^2 \theta + \nonumber \\
& & D \sin^4 \theta) \nonumber \\
\end{eqnarray}

\noindent and independently of $\theta$, these expressions become equal to 1 just if:

\begin{eqnarray} \label{fmixsit2}
A=2, G=4, D=2 
\end{eqnarray}

In expressions of before subsections we find that $A$ is related with $b_+, t_0$ and $s$ showing a periodic behavior for the two first, but decreasing until one half ($A \approx 1$) for large values of $s$ . A similar behavior is observed in $D$ in relation with $s$, but more centered in $j$ and $t_0$ parameters. Finally, $G$ exhibits diverse dependence of the parameters, but invariably note that $G \rightarrow 0$ if $s \rightarrow \infty$, which is responsible of (\ref{lim1infty}) and (\ref{lim2infty}).

\section{conclusions}
Study of entangled pure states distorted by Ising interaction doubted to parasite magnetic field shows that it's possible reach very good fidelities in the process of control. Nevertheless these schemes assume that repreparation is completely faithful. With direct and simple schemes of control (applying extra magnetic fields on site or separating particles and then apply secondary magnetic fields) just partial reconstruction could be reached, particularly for systems with $j \in \mathbb Q$. In addition, inhomogeneity of parasite magnetic field affect negatively this reconstruction (situation 2) because of formulas for $r, r'$ in (\ref{parameterssit1}).

Future works should be directed on measure, repreparation and feedback for optimization as was presented for one single qubit works \cite{branczyk1, xi1} (by example use of weak non-destructive measurements)  and considering other types of mixed states as presented here, combining both focuses for improve the fidelity. Alternatively, cross magnetic fields should be considered to study their effect in the repreparation \cite{branczyk1}.

In addition, correlation of these operations with the amount of entanglement should be considered in order to stablish some relations which suggest optimization of the repreaparation process as was suggested in \cite{delgado3}.

\section*{Aknowledgements}
I gratefully acknowledge to Dr. Sergio Martinez-Casas about some fruitful discussions about use of Ising model in quantum cellular automatas and to M. Sc. Jose Luis Gomez-Mu\~noz who developed the QUANTUM package based on {\it Mathematica} and who works in some Add-Ons and improvements for the package to make easier the review of some large calculations using Dirac notation involved in this work. 

\section*{FIGURE CAPTIONS}
\subsection*{Figure 1}
Schematic description of process of control. First, a system generates with equal probabilities, two non equivalent entangled states known by the observer but without any knowldge upon which was the state created at this time. After, during certain time, magnetic interaction between parts introduces an internal distorion on the original states. After of that, the observer would to correct this distortion and to try of recovering the original state by introducing approriate measurement of two parts of the distorted state. \\

\subsection*{Figure 2}
Behavior of function $f(B_-t,2Jt)=\phi-\phi'-4 J t$. Contour lines shown the corresponding values to: $-8\pi,-6\pi,...,6\pi,8\pi$ which are some of solutions where (\ref{cond1}) is fulfill simultaneously. Note that if $B_-/2J$ is fixed, $t$ could be selected in infinite ways still. 

\subsection*{Figure 3}
Average fidelity for three procedures of control as function of $\theta$ and $b_+$. Taking $j=\frac{1}{6}$ and $t=\frac{\pi}{2}$ as illustrative values (larger $t$ indtroduce more oscillations and larger $j$ "shortens" the distance of $F_N$ to 1 in agreement with (\ref{lim3})). {\it a)} Comparative aspect of $F_{DR_1}$ (black), $F_N$ (gray) and $F_{A,B}$ (white). Note specially the results (\ref{lim1})-(\ref{lim3}). {\it b)} Values of fidelity for $F_{SO}$; white dashed line shows the separation between two different procedures of control.

\subsection*{Figure 4}
Taking $j=\frac{1}{6}$ and $t=\frac{\pi}{2}$ for $\Re =[0,\frac{\pi}{2}] \times [0,5]$, $F_{DR_2}$ is compared with $F_{SO}$. {\it a)} Comparative aspect of $F_{SO}$ (black) below of $F_{DR_2}$ (chess boxed layer) exhibiting close similitudes except for the regions near of $b_+=\frac{(2n+1)}{2t}\pi, n \in \mathbb Z$. {\it b)} Values of fidelity for $F_{DR_2}$ placed between $\frac{1}{2}$ and $1$, but normally with values up of 0.8 (except in the three darker areas).

\subsection*{Figure 5}
Comparison between $F_N$ (black), $F_1$ (gray) and $F_2$ (white) for mixed states depending on $\theta$ and $s$, generated when there are uncertainty in the time of distortion. Taking $j=\frac{1}{6}$ and $b_+=1$ fixed, for {\it a)} $t_0=\frac{\pi}{2}$, {\it b)} $t_0=\frac{3 \pi}{4}$ and {\it c)} $t_0=\frac{7 \pi}{4}$, showing the complex behavior of control schemes upon the parameters.

\small  

\end{document}